\documentclass[%
 reprint,
superscriptaddress,
 amsmath,amssymb,
 aps,
]{revtex4-1}

\usepackage{color}
\usepackage{graphicx}
\usepackage{dcolumn}
\usepackage{bm}

\begin{document}

\preprint{APS/123-QED}

\title{Phonon mode softening and elastic properties of hafnium under pressure}

\author{Novoselov D.}
\email{novoselov@imp.uran.ru}
\author{Anisimov V.I.}
\author{Ponosov Yu.S.}
\affiliation{M.N. Mikheev Institute of Metal Physics, Ural Branch of the Russian Academy of Sciences, 18 S. Kovalevskaya Street, Yekaterinburg, 620137 Russian}
\affiliation{Ural Federal University, 19 Mira Street, Yekaterinburg, 620002 Russia}

\date{\today}

\begin{abstract}
The effect of pressure on the zone-center optical phonon modes and the elastic properties of hafnium has been studied by both experimental (Raman scattering) and theoretical (DFT) approaches. It was found an anomaly dependence of the phonon frequency of E$_{2g}$ mode in the pressure range from 0 to 67 GPa.
The calculated electronic structure of hafnium shows significant changes under pressure, which have a pronounced spatial anisotropy similar to the anisotropy of the observed phonon properties.
The dependencies of the elastic properties on pressure including the components of the elasticity tensor, bulk and shear modulus, Poisson ratio and Universal anisotropy index obtained during the calculations has characteristic features indicating the occurrence of the structural phase transition.

\begin{description}
\item[PACS numbers]
\end{description}
\end{abstract}

\pacs{Valid PACS appear here}
\maketitle


\section{Introduction}

Technologically important transition metals of group IV (Ti, Zr, Hf) possess a hexagonal close-packed structure (hcp-$\alpha$) at atmospheric pressure. When the pressure increases, they all undergo successive structural transitions, initially to the $\omega$ phase with a simple hexagonal lattice with three atoms per unit cell, and then to the cubic phase $\beta$ with a bcc lattice. The crystal structure of hafnium under pressure was investigated by X-ray diffraction in \cite{Xia1990} where it was reported that the transition to the $\omega$ phase began at pressure of 45.8 GPa and one was completed at 58.3 GPa. 
The reverse transition from $\omega$ phase had a significant hysteresis, starting at 30 GPa and ending at 19.3 GPa. Subsequent X-ray investigation of hafnium at high pressures \cite{Hrubiak2012} did not reveal a transition to the $\omega$ phase at room temperature in the pressure region up to 51 GPa. A mixture of $\alpha$ and $\omega$ phases was observed only with an increase in temperature up to 770 K at a pressure of 45 GPa. The authors explained the discrepancy obtained by the effects of nonhydrostaticity in experiment \cite{Xia1990} which can lower the temperature of the martensitic transition. Another X-ray study \cite{Pandey2014a} at room temperature confirmed the existence of a sequence of transformations $\alpha \rightarrow \omega \rightarrow \beta$ in pure Hf. In \cite{Pandey2014a} $\alpha \rightarrow \omega$ transition with a wide hysteresis began above the pressure P = 50.6 GPa, and the $\omega \rightarrow \beta$ transition occurred above the pressure P = 65 GPa. It has been shown that the use of different pressure transmitting media results in small shifts in the transition pressures $\alpha \rightarrow \omega$, but the transition itself can be suppressed when the impurity concentration in the sample increases to 7.5\%.
In spite of the fact that some \textit{ab initio} calculations of transition pressures have been carried out \cite{Gyanchandani1990,Ahuja1993,Ostanin2000,Jomard1998,Fang2011,Hao2011a,Qi2016} the mechanism of structural instability of hafnium has been studied insufficiently both experimentally and theoretically. A recent experimental and theoretical study of the elastic properties of hafnium \cite{Qi2016} showed the presence of softening of shear moduli with an increase in pressure above 10 GPa which could indicate the precursors of the $\alpha \rightarrow \omega$ transition.

Raman spectroscopy can provide information not only about the structural state of a substance under pressure but also about the effect of changes in interatomic distances on the vibrational and electronic spectrum. Thus, the Raman studies of titanium and zirconium \cite{Olijnyk1997,Ponosov2012} revealed an anomalous behavior of the frequency of the E$_{2g}$ phonon which is softened in the $\alpha$ phase under pressure. The relationship between the E$_{2g}$ phonon frequency and the elastic modulus $C_{44}$ suggested a decrease in the latter under pressure which was confirmed by subsequent ultrasonic measurements \cite{Liu2008}. 
For hafnium, Raman measurements, as well as calculations of the phonon spectrum, have not yet been reported.

Despite the substantial interest in this system from the scientific community to date, there is a lack of unambiguous insight on the relationship between phonon properties and structural stability at high pressures in hafnium.
In our work, the evolution of the phonon spectrum of hafnium in the pressure region 0-67 GPa at room temperature is investigated by Raman spectroscopy. For the analysis and interpretation of experimental data, first-principles calculations of phonon frequencies and elastic properties at different pressures have been performed by using DFT.

\section{Methods}
\subsection{Experiment}
Samples for measurements were cut from a hafnium single crystal with a purity of 99.8\% having an residual resistivity ratio of $\rho_{300K} / \rho_{4.2K} = 27 $. Thin (20 $\mu$m) plates with transverse dimensions up to 100 $\mu$m were placed in the center of the flat culets (300 $\mu$m in diameter) of diamond anvils. The experiment was carried out without the use of a pressure transmitting medium: the sample was compressed between the anvil and the gasket. The pressure in the cell was determined from the shift of the high-frequency edge of the T$_{2g}$ phonon diamond line in the sample region under study. Preliminary calibration of the dependence of the diamond line frequency on the pressure in the used anvils was carried out using ruby luminescence method: the result practically coincided with the data of \cite{Akahama2006}. Two runs were performed. In the first, a gasket was made of steel, and the highest pressure was 50 GPa. In the second run, the pressure reached 67 GPa, and the use of a rhenium gasket provided an additional possibility of controlling the pressure \cite{Goncharov2002}. The spectra were excited by lines of solid state (532 nm) and helium-neon (633 nm) lasers focused (50 objective) to a spot of 5-10 $\mu$m in diameter. The scattered light was recorded in backscattering geometry using a Renishaw RM1000 spectrometer equipped with edge filters to exclude low-frequency Rayleigh scattering with a cut-off $\simeq$ 40 cm$^{-l}$ and a cooled multichannel CCD detector.

\subsection{Calculations}
Crystal structure data for Hf at different values of pressure were taken from the paper \cite{Pandey2014} and one has been used as an initial data for \textit{ab-initio} electronic structure calculations based on the GGA approximation of the DFT implemented in VASP package \cite{Kresse1996,Blochl1994,Perdew1996}. 
In all the calculations the valence configuration is 5p6s5d with four valence electrons.
The frozen-phonon method has been used to evaluate the optical phonon mode frequencies at the $\Gamma$-point of the Brillouin zone. 
The plane wave energy cutoff was set to 450 eV and we have used $\Gamma$-centered k-point mesh 12x12x12 for pressure values from 0 up to 65 GPa. Self-consistent convergence of the energy was set to 10$^{-8}$ eV/cell and the force convergent criteria of 10$^{-7}$ eV/\AA.

\section{Results and discussion}

\begin{figure}[!ht]
\includegraphics[width=0.35\textwidth]{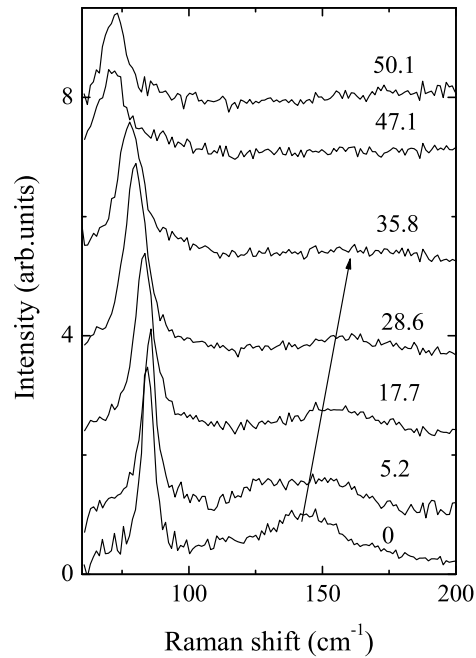}
\caption{Raman spectra of Hf at different pressures measured in the first run using 633 nm excitation wavelength.}
\label{fig1}
\end{figure}

In the hcp phase of Hf, there are three optical phonon modes: B$_{1g}$ and doubly degenerate E$_{2g}$ modes. The first one corresponds to the atomic displacements of the Hf ions along the z-direction (the green arrows in the \ref{charge}). 
Whereas the E$_{2g}$ phonon mode associated with vibrational motion in the xy-plane  (the red and blue arrows in the \ref{charge}).
According to the selection rules for hcp crystals in the Raman spectrum of the first order, one line - the E$_{2g}$ optical phonon at the $\Gamma$ point of the Brillouin zone should be observed. 
In the spectra of hafnium at P=0 GPa (figure \ref{fig1}) very narrow (about 2-3 cm$^{-1}$) peak at the frequency of 85 cm$^{-1}$ is recorded, whose energy is close to the value of the energy of the TO $\left[0001\right]$ mode at $\Gamma$ obtained in the neutron experiment \cite{Stassis1981}.

In addition to this peak, broad lines in the frequency range of 145 and 230-290 cm$^{-1}$ were found in the spectra. 
They are observed only for a parallel orientation of the polarization of the incident and scattered light (A$_{1g}$ symmetry), which agrees with the selection rules for the two-phonon spectra activity in hcp crystal. 
A comparison with the phonon dispersion from the neutron experiment \cite{Stassis1981} suggests that the feature near 145 cm$^{-1}$ is definitely associated with overtone scattering from transverse acoustic phonons at the M and A points of the Brillouin zone, while in the 235-290 cm$^{-1}$ region there are overtone excitations from several rather flat longitudinal and transverse branches near the point M. 
A rather large intensity of two-phonon scattering makes it possible to obtain information on the density of phonon states in hafnium from the Raman experiment.
The spectra obtained under pressure in the first run (figure \ref{fig1}) show a slight increase in the E$_{2g}$ mode frequency with an increase in pressure up to 10-15 GPa. 
With further increase in pressure, an anomalous softening of the E$_{2g}$ mode frequency is observed, while the energy of the broad feature at 145 cm$^{-1}$ is increased by 15\% with the growth of P up to 35 GPa (figure \ref{fig1}). 

\begin{figure}[!ht]
\includegraphics[width=0.45\textwidth]{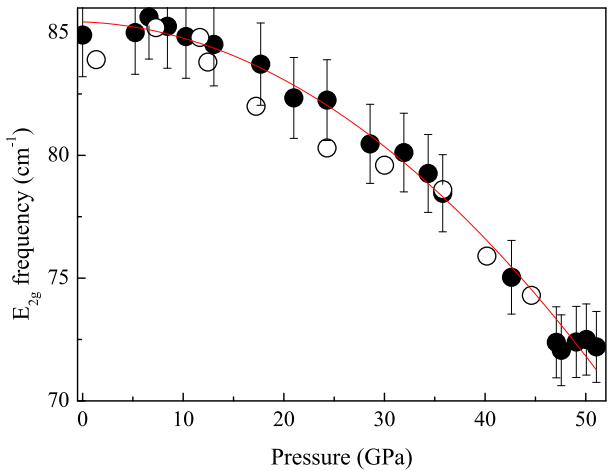}
\caption{E$_{2g}$ phonon mode frequency of Hf vs different pressure  from the first run depicted by filed circles (with increase a pressure), empty circles (with decrease a pressure) and red line (approximation).}
\label{fig2}
\end{figure}

At higher pressures, the intensity of this feature is lower than the observation limit. The high-frequency peak at 230-290 cm$^{-1}$ was not observed in measurements in the cell. 
The dependence of the E$_{2g}$ mode frequency vs pressure from the first run is shown in figure \ref{fig2}. 
Its energy drops by $\sim$17\% to 50 GPa. 
Such a softening of the mode was previously observed in the $\alpha$-phase of zirconium \cite{Olijnyk1997} and was supposed to be due to changes in the electronic structure under pressure.
Despite a significant decrease in the intensity of the E$_{2g}$ mode to 50 GPa in the first run, no Raman-active E$_{2g}$ phonon of the $\omega$ phase was observed in the spectra. 
On pressure release, we did not observe hysteresis in the behavior of the phonon frequency (see figure \ref{fig2}).

\begin{figure}[!ht]
\includegraphics[width=0.35\textwidth]{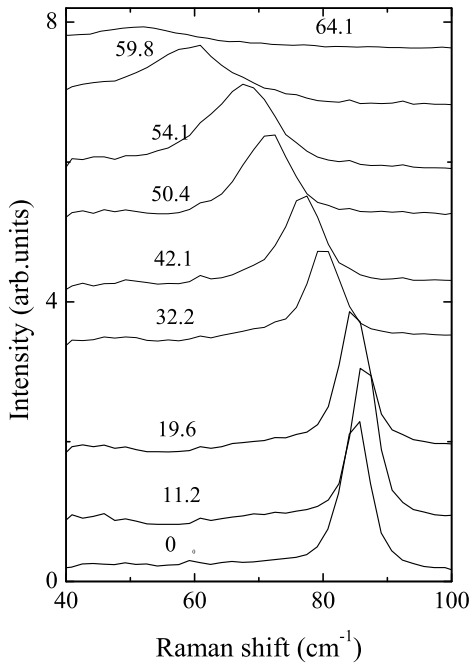}
\caption{Raman spectra of Hf at different pressures measured in the second run using 532 nm excitation wavelength.}
\label{fig3}
\end{figure}

\begin{figure}[!ht]
\includegraphics[width=0.45\textwidth]{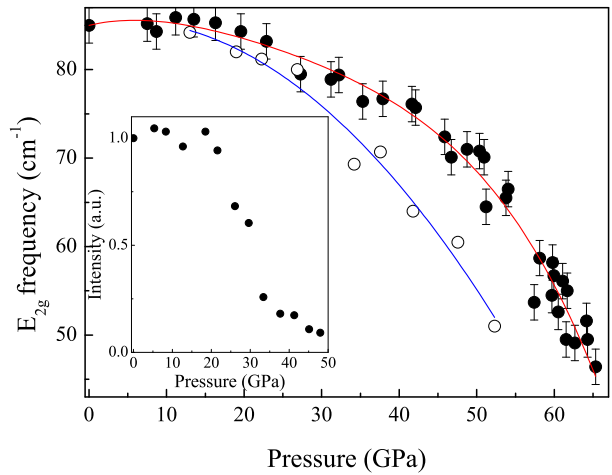}
\caption{E$_{2g}$ phonon mode frequency of Hf vs pressure from the second run depicted by filed circles (with increase a pressure), empty circles (with decrease a pressure), red and blue lines (approximation). The dependence of intensity under pressure release is shown on the inset.}
\label{fig4}
\end{figure}

Raman spectra of Hf measured in the second run up to 67 GPa are shown in figure \ref{fig3}.
The pressure dependence of the E$_{2g}$ frequency in the second run (see figure \ref{fig3} and \ref{fig4}) coincides with the first run data in the region up to 50 GPa.
With a further increase in pressure, the frequency and intensity of the E$_{2g}$ mode decrease significantly. 
The highest pressure at which this line was observed was 64 GPa. 
With an increase in pressure above 50 GPa, the reflection of the investigated region of the sample in the microscope changed its intensity and became shiny at the highest pressure of 67 GPa.
At this pressure, we could not observe the phonon line which implies the achievement of transition pressure. 
On pressure release, the line appeared again at a pressure of 52 GPa, and its frequency showed a significant hysteresis, recovering at pressures $<$20 GPa (figure \ref{fig4}). 
We also observed that the sample changed appearance from shiny to black in reflected light on pressure release to 50 GPa.
We believe that the sharp drop in the frequency of the E$_{2g}$ mode above 50 GPa is associated with the onset of the transition to the $\omega$ phase (as in study \cite{Pandey2014a}). 
However, in our measurements, it was not possible to observe the Raman-active E$_{2g}$ phonon of the hexagonal high-pressure phase, either on loading or when unloading the sample. 
Such a line was well observed in the Raman spectra of the $\omega$ phase of titanium and zirconium \cite{Olijnyk1997,Ponosov2012}. 
The reason for that may be its weak intensity or its low frequency unattainable for our spectrometer ($<$ 45 cm$^{-1}$). 
The frequency hysteresis of the E$_{2g}$ mode in the 20-50 GPa range (figure \ref{fig4}), which was not observed in the first experiment up to 50 GPa (figure \ref{fig2}), implies the appearance of phase in the second run.
Besides, the intensity of the phonon line increased significantly on pressure release in the region 40-20 GPa (inset in figure \ref{fig4}), which agrees with the data of \cite{Pandey2014a} and perhaps indicates an increase in the amount of phase. 
These facts imply that, in spite of the absence of the $\omega$ phase line in our spectra, the $\alpha$ and $\omega$ phases coexist in the region of 50-64 GPa on pressure increase and in the region of 52-20 GPa on pressure release.
Disappearance at the highest pressure, possibly, implies a transition to the $\beta$ phase. 
Another explanation the absence of the $\omega$ phase line in our spectra assumes a direct transition of the $\alpha$ phase to the $\beta$ phase.

Although Raman experiment probes only $\Gamma$=0 phonon, it is obvious that at least the transverse optical phonons with q = 1/6 $\left[ 010 \right]$ in hafnium will be significantly softened under pressure. A significant decrease of the elastic modulus $C_{44}$ (coupled with the E$_{2g}$ phonon frequency) indicates the involvement of acoustic phonons with the same wave vector. These observations support the transformation mechanism proposed in \cite{Ranganathan2008}.

\begin{figure}[!ht]
\includegraphics[width=0.5\textwidth]{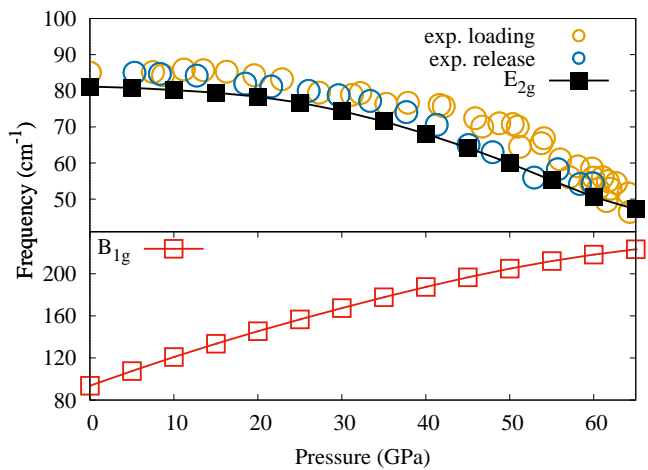}
\caption{Calculated frequencies of E$_{2g}$ (black curve with filled squares in the upper panel) and B$_{1g}$ (red curve with empty squares in the lower panel) optical phonon modes at $\Gamma$-point and the experimental data indicated by blue and orange empty circles correspond to the decrease and increase pressure.}
\label{phonons}
\end{figure}

To clarify the relationship of the anomalous behavior of the phonon mode with electronic degrees of freedom of the system under the study, the calculations in the framework of the density functional theory were carried out for $\alpha$-Hf crystal structures at different pressures.
During the GGA calculations, the values of the phonon mode frequencies at the $\Gamma$-point were obtained by the frozen-phonon method in the pressure range from 0 up to 65 GPa. 
Our calculations confirmed the main results obtained in the experiment (see figure \ref{phonons}). 
As we can see from the upper panel of the figure the frequency of E$_{2g}$ mode is softening which is counterintuitive behavior during compression. 
At the same time, the calculated frequency of B$_{1g}$ phonon mode increases under pressure (lower panel in figure \ref{phonons}).
The small difference between the calculated and experimental data may be because non-hydrostatic pressure is applied in the experiment.
Also, it is possible \cite{Fang2011} that taking into account in the calculations the spin-orbit interaction can slightly improve the quantitative agreement with experiment.

\begin{figure}[!ht]
\includegraphics[width=0.5\textwidth]{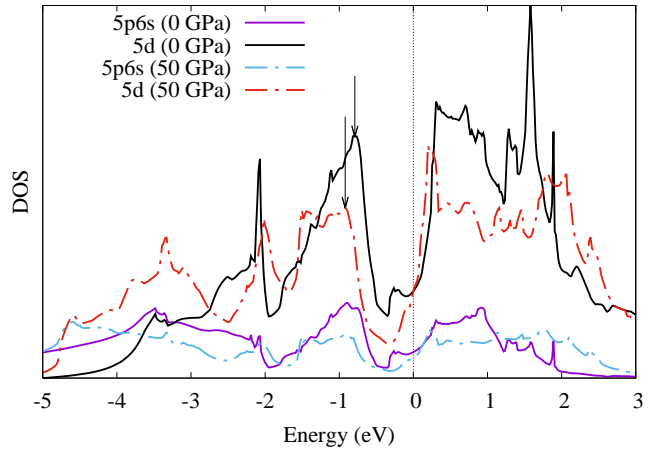}
\caption{The density of 5p6s- (violet and blue) and 5d-Hf (black and red) states at 0 and 50 GPa pressure is depicted by solid and bar-doted curves correspondingly. 
The Fermi level is placed at zero energy and denoted by the dashed vertical line. 
The vertical arrows indicate the centers of the d-peak near the Fermi in the valence band.}
\label{total_dos}
\end{figure}

\begin{figure}[!ht]
\includegraphics[width=0.5\textwidth]{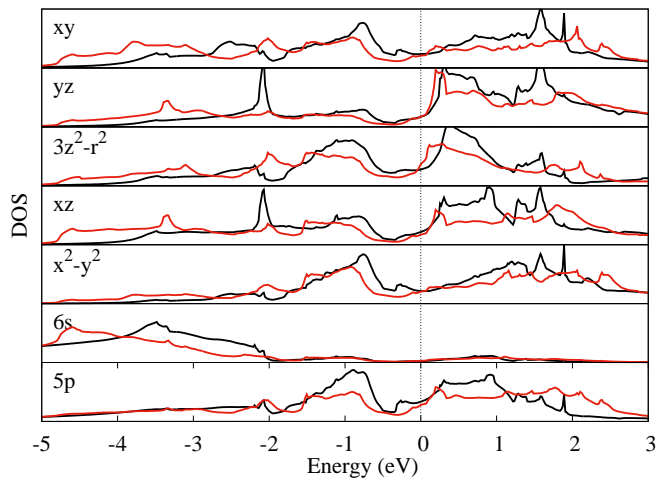}
\caption{The density of states for 5d and 6s, 5p levels at 0 GPa (black curve) and 50 GPa (red curve) pressures. The Fermi level corresponds to zero energy.}
\label{partial_dos}
\end{figure}

\begin{figure}[!ht]
\begin{minipage}[h]{0.85\linewidth}
\center{\includegraphics[width=1\textwidth]{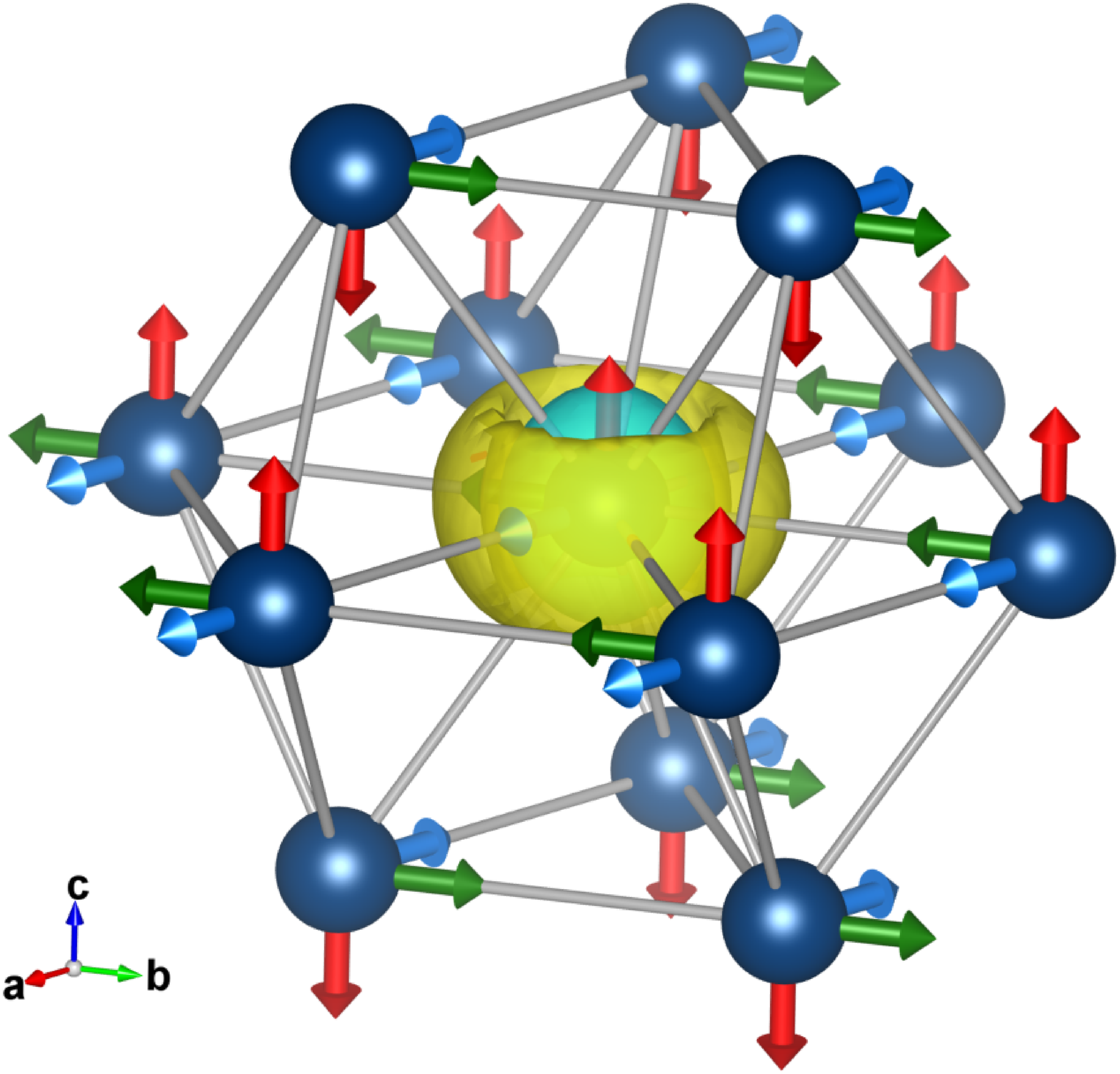}} \\(a)
\end{minipage}
\vfill
\begin{minipage}[h]{0.9\linewidth}
\center{\includegraphics[width=1.0\textwidth]{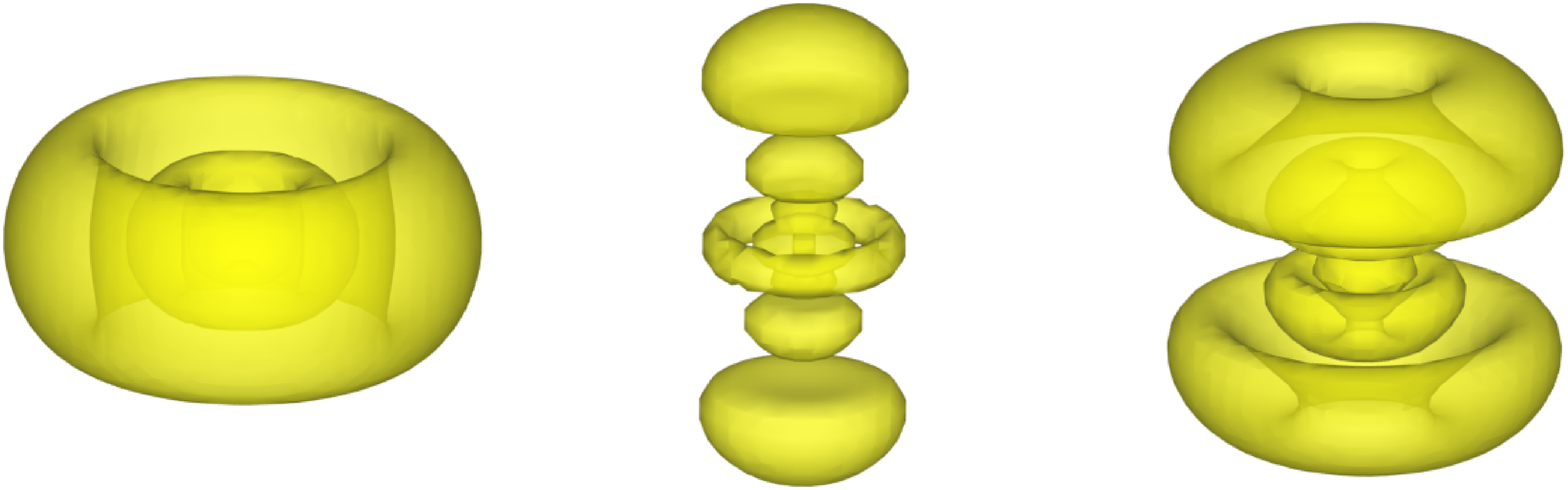}} \\(b)
\end{minipage}
\caption{(a) The crystal structure of $\alpha$-Hf and the isosurface of the charge density difference (showed for the central atom only) under compression. The navy blue spheres denote the hafnium atoms. The yellow and blue regions of the isosurface indicate positive and negative differences in charge densities correspondingly. 
The arrows denote the directions of the atomic displacements corresponding the phonon vibrational modes: green and blue - double degenerate E$_{2g}$ and red - B$_{1g}$.
(b) The spatial shape of the atomic 5d-orbitals obtained as the squared moduli of the 5d wave functions (from left to right): sum of x$^2$+y$^2$ and xy, 3z$^2$+r$^2$, sum of xz and yz.}
\label{charge}
\end{figure}

To determine the influence of pressure on the electronic structure of the $\alpha$ phase of hafnium, the total and partial densities of states were obtained. 
The density of hafnium 5p6s and 5d states are presented in figure \ref{total_dos} for two pressure values 0 and 50 GPa. 
The general shape and positions of the peaks are in good agreement with previously published data \cite{Ahuja1993,Fang2011}.
There are two competing effects which influence on the occupation of the d-level near the Fermi energy with a change in pressure. First of them connected with the effect of broadening of the relatively narrow peak under compression which leads to decrease of the d-state occupation due to the rise of the area under the curve above the Fermi level. The second one is due to the shift of the peak downward in energy, which leads to an increase the occupation of the corresponding states.
In figure \ref{total_dos} the vertical arrows indicate the centers of the peaks under the Fermi level. As we can see, there is a shift of the peak to lower energy when pressure is applied.
The value of the shift is about 0.13 eV.
Along with this effect, the broadening of the bands below -2 eV towards lower energies is observed.
Besides, it can be observed that the density of 5d-states at the Fermi level remain unchanged under compression as if the band would be attached to the zero energy.
Such redistribution of the density of states leads to increase of the occupation of the 5d level.
The figure \ref{total_dos} also shows the density of sp-Hf states. The position and the structure of the peaks on the curves quite clearly repeat ones in the d-band which confirms the presence of a strong hybridization between corresponding states.
The growth of the number of d-electrons is associated with the rise of the hybridization between d- and sp-levels of neighboring atoms with increasing the pressure due to decreasing of the interatomic distances which leads to reducing the occupation of sp-states. In the $\alpha$-phase the distances between the nearest neighboring hafnium atoms are different for atoms in the $ab$-plane and atoms in neighboring planes. With increasing pressure, these distances and also the difference between them decrease up to a pressure of 70 GPa, at which the corresponding lengths coincide \cite{Pandey2014}.
The number of 5d-electrons is changing from 2.11 at 0 GPa pressure to 2.30 at 50 GPa which agrees well with the previously published data \cite{Fang2011}. 
For a more detailed analysis of the change in the density of states with increasing pressure the partial density of 5d-Hf states were constructed and are shown in figure \ref{partial_dos} in comparison with 6s- and 5p-Hf states.
The figure \ref{partial_dos} shows clearly that 6s states have the largest hybridization with 5d states with xy and yz symmetry in the bottom of the whole density of states. 
While 5p states hybridize predominantly with x$^2$-y$^2$, xy and 3z$^2$-r$^2$ states in the region from -2.5 eV up to the Fermi level as indicated by a pronounced peak near -0.8 eV.
From figure \ref{partial_dos}, it follows that the occupation of 6s states remains unchanged whereas the area under 5p curve below E$_F$ is decreased when pressure is applying.
As we can see from figure \ref{partial_dos} the orbitals with x$^2$-y$^2$ and xy symmetry undergo the most noticeable shift of that relatively narrow peak edge which may indicate their dominant contribution to the redistribution of the density of states near the Fermi level as a result of the growth of the pd-hybridization.
The average value of the peak edge shift has the value about 0.25 eV.
At the same time, the edge of the peak under the Fermi level on the density of states of the $3z^2-r^2$ orbital has no such shift with increasing pressure.
Thus, the redistribution of the density of states, which have the largest contribution near the Fermi level, has a pronounced spatial anisotropy.
These observations are consistent with the results of an analysis of the change in the charge density at external pressure.
In figure \ref{charge} (a) it is presented the schematically view of the crystal structure of $\alpha$-Hf and the isosurface of the charge density difference between 45 GPa and 65 GPa pressures as well as the spatial shapes of 5d atomic orbital combinations in \ref{charge} (b). 
The important thing here is the shape of the isosurface of the positive charge density difference which repeats the shape of the square moduli  sum of the wave functions of general 5d states with x$^2$+y$^2$ and xy symmetry (left shape in \ref{charge} (b)) in a greater degree than ones of 3z$^2$+r$^2$ or xz and yz symmetry (center and right shapes in \ref{charge} (b)). 
The negative difference of the charge density has a spherical-like shape which is probably due to a decrease in the occupation of the 5p-Hf states. 
This visually illustrates the effect of redistribution of the density of states which is connected with the spatial orientation of the partially filled d- and sp-orbitals and with increasing of the overlap area between them under pressure.
In this way, during the pressure rise, the charge density becomes advantageous to redistribute to the xy-plane but not in the z-direction.

\begin{figure}[!ht]
\includegraphics[width=0.5\textwidth]{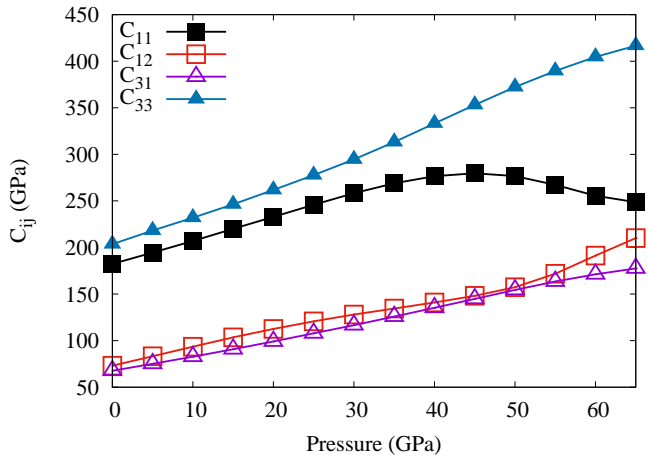}
\caption{The dependencies of the elastic tensor components C$_{ij}$ on pressure.}
\label{Cij}
\end{figure}

\begin{figure}[!ht]
\includegraphics[width=0.5\textwidth]{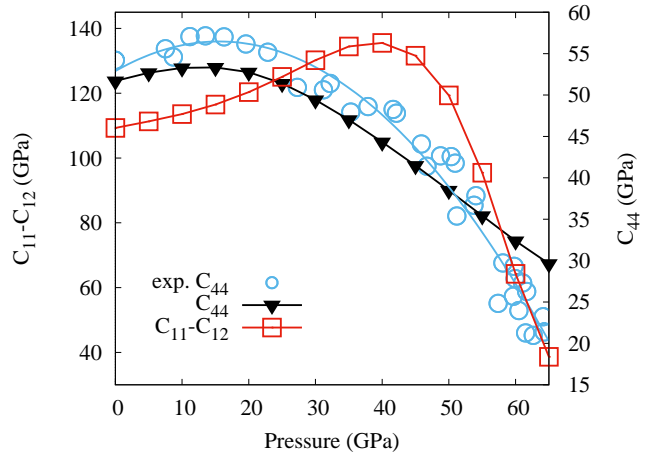}
\caption{The dependencies of the elastic tensor component difference $C_{11}$-$C_{12}$ as well as $C_{44}$ on pressure are denoted by red (left axis) and black curves (right axis) with empty squares and filed triangles correspondingly. The blue circles and blue curve correspond to the $C_{44}$ component of the elastic tensor calculated by the formula (\ref{C44_from_E2g}) from the experimental data for the frequency of E$_{2g}$ phonon mode. The blue curve data were obtained by the least square approximation.}
\label{C11-12}
\end{figure}

\begin{figure}[h]
\includegraphics[width=0.5\textwidth]{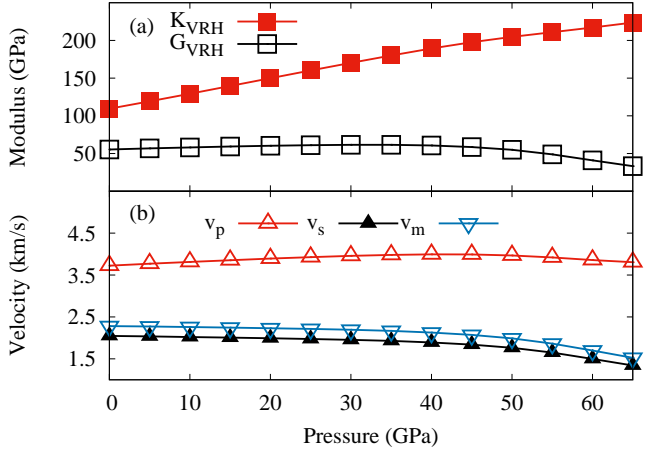}
\caption{(a) Adiabatic bulk module K and shear G module, (b) Calculated compressional wave $v_p$, shear wave $v_s$ and sound wave $v_m$ velocities.}
\label{multiplot2}
\end{figure}

\begin{figure}[h]
\includegraphics[width=0.5\textwidth]{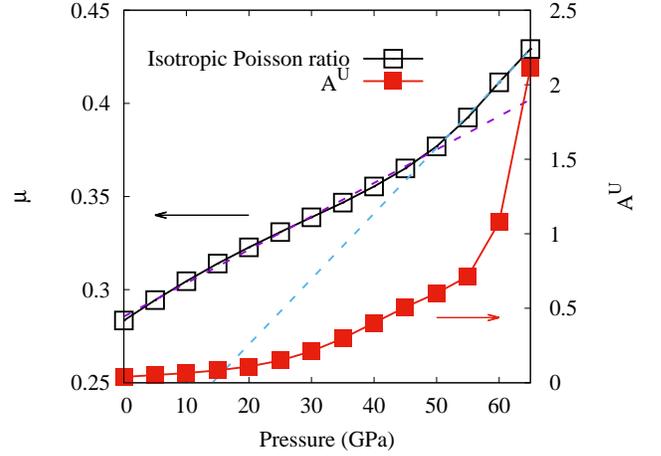}
\caption{Isotropic Poisson ratio $\mu$ and Universal anisotropic index A$^U$.}
\label{multiplot3}
\end{figure}

To understand a mechanical instability of the $\alpha$-Hf structure and possible causes of $\alpha \rightarrow \omega$ phase transition the change of the elastic and the mechanical properties under pressure were investigated.
The components of the elasticity tensor at various values of the external pressure were determined in the course of the calculations. Figure \ref{Cij} shows the dependencies of the values of the non-zero components of the elasticity tensor on pressure which values are in good agreement with those presented in the literature \cite{Qi2016}.
The figure \ref{C11-12} shows the results of calculating of the elastic constant C$_{44}$ obtained both during the DFT calculations and from the experimental data by using a simple model \cite{Verma1971,Olijnyk2002} of the lattice dynamics of hcp metals implementing the relationship between the value of the elastic constant $C_{44}$ and the phonon frequency of E$_{2g}$ mode in the following form:
\begin{eqnarray}
C_{44}=\frac{mc}{4\sqrt{3}a^2}\omega_{E_{2g}}^2,
\label{C44_from_E2g}
\end{eqnarray}
here $a$ and $c$ are the lattice constants, $m$ is the atomic mass, and $\omega_{E_{2g}}$ is a frequency of the E$_{2g}$ mode. 
The value of the C$_{44}$ elastic constant shows a dramatic fall ($\sim$ 3 times) at the highest pressure P=65 GPa.
In addition to condition $C_{44}>0$ which in this case is weakening, there is also one of the necessary and sufficient criteria for elastic stability of the hexagonal crystal system \cite{Mouhat2014}:
\begin{eqnarray}
C_{11}>|C_{12}|.
\end{eqnarray}
And as we can see from figure \ref{C11-12}, this condition is sharply weakened more than three times with increasing pressure above 40 GPa. Such behavior of the considered dependence explicitly indicates the structural instability of the crystal in the corresponding pressure range.
The $C_{11}(=C_{22})$ elastic constant characterized a longitudinal compression (Young's modulus) along the x(y)-direction has the change in the sign of the derivative in the pressure region of the phase transition. 
Simultaneously, the behavior of the $C_{33}$ (z-direction) elastic constant does not undergo similar changes and monotonically increases with increasing pressure. 
The difference in the dependence of these moduli on pressure indicates a change in the anisotropy of the crystal. The Universal elastic anisotropy index $A^U$ was calculated to verify this assertion by the definition proposed by the authors of the article \cite{Ranganathan2008} in the following view
\begin{eqnarray}
A^U=5\frac{G^V}{G^R}+\frac{K^V}{K^R}-6,
\end{eqnarray}
where $G^V$, $G^R$ and $K^V$, $K^R$ are the shear and bulk modulus' Voigt and the Reuss estimate respectively.
The dependence of Universal elastic anisotropy index on the pressure is presented in figure \ref{multiplot3}. As we can see the value of $A^U$ monotonically increases with increasing the pressure, and it undergoes a jump in the region above 45 GPa which may indicate the tendency to the transition into a more anisotropic phase.
The figure \ref{Cij} also shows the curves of the dependences of the $C_{12}$ and $C_{31}$ elastic constants on pressure which are defined by a transverse extension. As can be seen, the latter has an almost linear dependence on pressure, while the former has an extremum in the region of 45 GPa. This effect is also associated with a slight change in the slope of the Poisson's ratio curve versus pressure at 45 GPa which can be seen in figure \ref{multiplot2} (c).
Thus, there is a difference between the elastic properties in z- and x,y-directions. The last one show an anomalous behavior which may serve as a precursor of the structural instability occurrence and the phase transition to a more stable phase. 

On the basis of the values of the elasticity tensor components, the elastic moduli, the velocities of elastic waves and also the sound velocity were determined in the pressure range from 0 to 65 GPa.
The averaged bulk $K_{VRH}$ and shear $G_{VRH}$ modulus were calculated by the Voigt-Reuss-Hill average scheme \cite{Hill1952}. Obtained dependencies of $K_{VRH}$ and $G_{VRH}$ on pressure is presented in figure \ref{multiplot2} (a).
The negative slope of the shear modulus pressure dependence has been found above 40 GPa which may indicate the shear instability of $\alpha$-Hf preceding the transition as was established for $\alpha$-Zr \cite{Liu2008}. 
The compressional $v_p$ and shear $v_m$ waves velocity average values has been evaluated by the formula \cite{Hao2011}
\begin{eqnarray}
v_p=\sqrt[]{\frac{K_{VRH}+\frac{4}{3}G_{VRH}}{\rho}}, v_s=\sqrt[]{\frac{G_{VRH}}{\rho}},
\label{comp_shear_velocity}
\end{eqnarray}
where $\rho$ is the density. 
The average value of the sound velocity $v_m$ \cite{Hao2011} was calculated as:
\begin{eqnarray}
v_m=\left[1/3\left(\frac{2}{v_s^3}+\frac{1}{v_p^3}\right)\right]^{-1/3}.
\label{sound_velocity}
\end{eqnarray}
The pressure dependencies of the propagation velocities of elastic waves are shown in figure \ref{multiplot2} (b). As can be seen from this figure, the monotonic growth of the velocities in the pressure region up to 20 GPa is replaced by a monotonic decrease up to 65 GPa. This effect is most clearly seen for the shear wave velocity and the sound velocity. Such behavior may indicate a structural instability of the crystal lattice at pressures above 20 GPa.

To estimate the Isotropic Poisson ratio $\mu$ one was used the following formula \cite{DeJong2015}:
\begin{eqnarray}
\mu = \frac{3K_{VRH}-2G_{VRH}}{6K_{VRH}+2G_{VRH}}.
\label{poisson}
\end{eqnarray}
The dependence of the $\mu$ on pressure is shown in figure \ref{multiplot3} by the solid black curve with empty squares.
In the same figure, two dashed lines (violet and blue) are also given, which were obtained by the least squares approximation method in two corresponding pressure ranges from 0 to 35 GPa and from 50 to 65 GPa to illustrate the presence of two regions with different slopes of the Poisson's ratio curve versus pressure as well as existence of wide transition region between them.

\section{Conclusion}
In the present study, the Raman scattering experiment under pressure found the phonon softening the effect of the E$_{2g}$ mode corresponding to the oscillations of Hf-atoms in the xy-plane. Theoretical calculations of an electronic structure carried out within the framework of the DFT also confirm the presence of this effect.
The observed phonon mode softening effect under pressure reflects the close relationship between the electronic and lattice degrees of freedom in $\alpha$-Hf.
It was found that two effects associated with the charge distribution occur simultaneously with increasing pressure. Growth in the external pressure leads to the rise of 5d-Hf shell occupation due to sp-d hybridization strengthening. Furthermore, there is a redistribution of the electron density between 5d sub-shells in favor of sub-shells corresponding to the orbitals oriented in the xy-plane.
The values of the components of the elasticity tensor, the bulk and shear moduli, the Isotropic Poisson ratio, and the Universal anisotropy index obtained during the calculations demonstrate unusual behavior with increasing pressure that indicates the structural instability of the alpha phase and can serve as a good indicator of the appearance of the phase transition.
It is known that the temperature of the superconducting transition in hafnium increases substantially under pressure. T$_c$ is 0.128 K at atmospheric pressure and increases to 8K at 64 GPa \cite{Bashkin2004}. Anomalous softening of the E2g mode can be one of the causes of Tc growth due to a change in the electron-phonon interaction.

\section{Acknowledgements}
The present work was supported by the grant of the Russian Science Foundation (project no. 14-22-00004). The authors thank Patrakov E.I. for electron probe microanalysis of the samples as well as Streltsov S.V. and Korotin Dm.M. for useful discussion.

\bibliography{Mendeley.bib}

\begin{thebibliography}{28}%
\makeatletter
\providecommand \@ifxundefined [1]{%
 \@ifx{#1\undefined}
}%
\providecommand \@ifnum [1]{%
 \ifnum #1\expandafter \@firstoftwo
 \else \expandafter \@secondoftwo
 \fi
}%
\providecommand \@ifx [1]{%
 \ifx #1\expandafter \@firstoftwo
 \else \expandafter \@secondoftwo
 \fi
}%
\providecommand \natexlab [1]{#1}%
\providecommand \enquote  [1]{``#1''}%
\providecommand \bibnamefont  [1]{#1}%
\providecommand \bibfnamefont [1]{#1}%
\providecommand \citenamefont [1]{#1}%
\providecommand \href@noop [0]{\@secondoftwo}%
\providecommand \href [0]{\begingroup \@sanitize@url \@href}%
\providecommand \@href[1]{\@@startlink{#1}\@@href}%
\providecommand \@@href[1]{\endgroup#1\@@endlink}%
\providecommand \@sanitize@url [0]{\catcode `\\12\catcode `\$12\catcode
  `\&12\catcode `\#12\catcode `\^12\catcode `\_12\catcode `\%12\relax}%
\providecommand \@@startlink[1]{}%
\providecommand \@@endlink[0]{}%
\providecommand \url  [0]{\begingroup\@sanitize@url \@url }%
\providecommand \@url [1]{\endgroup\@href {#1}{\urlprefix }}%
\providecommand \urlprefix  [0]{URL }%
\providecommand \Eprint [0]{\href }%
\providecommand \doibase [0]{http://dx.doi.org/}%
\providecommand \selectlanguage [0]{\@gobble}%
\providecommand \bibinfo  [0]{\@secondoftwo}%
\providecommand \bibfield  [0]{\@secondoftwo}%
\providecommand \translation [1]{[#1]}%
\providecommand \BibitemOpen [0]{}%
\providecommand \bibitemStop [0]{}%
\providecommand \bibitemNoStop [0]{.\EOS\space}%
\providecommand \EOS [0]{\spacefactor3000\relax}%
\providecommand \BibitemShut  [1]{\csname bibitem#1\endcsname}%
\let\auto@bib@innerbib\@empty
\bibitem [{\citenamefont {Xia}\ \emph {et~al.}(1990)\citenamefont {Xia},
  \citenamefont {Parthasarathy}, \citenamefont {Luo}, \citenamefont {Vohra},\
  and\ \citenamefont {Ruoff}}]{Xia1990}%
  \BibitemOpen
  \bibfield  {author} {\bibinfo {author} {\bibfnamefont {H.}~\bibnamefont
  {Xia}}, \bibinfo {author} {\bibfnamefont {G.}~\bibnamefont {Parthasarathy}},
  \bibinfo {author} {\bibfnamefont {H.}~\bibnamefont {Luo}}, \bibinfo {author}
  {\bibfnamefont {Y.~K.}\ \bibnamefont {Vohra}}, \ and\ \bibinfo {author}
  {\bibfnamefont {A.~L.}\ \bibnamefont {Ruoff}},\ }\href {\doibase
  10.1103/PhysRevB.42.6736} {\bibfield  {journal} {\bibinfo  {journal}
  {Physical Review B}\ }\textbf {\bibinfo {volume} {42}},\ \bibinfo {pages}
  {6736} (\bibinfo {year} {1990})}\BibitemShut {NoStop}%
\bibitem [{\citenamefont {Hrubiak}\ \emph {et~al.}(2012)\citenamefont
  {Hrubiak}, \citenamefont {Drozd}, \citenamefont {Karbasi},\ and\
  \citenamefont {Saxena}}]{Hrubiak2012}%
  \BibitemOpen
  \bibfield  {author} {\bibinfo {author} {\bibfnamefont {R.}~\bibnamefont
  {Hrubiak}}, \bibinfo {author} {\bibfnamefont {V.}~\bibnamefont {Drozd}},
  \bibinfo {author} {\bibfnamefont {A.}~\bibnamefont {Karbasi}}, \ and\
  \bibinfo {author} {\bibfnamefont {S.~K.}\ \bibnamefont {Saxena}},\ }\href
  {\doibase 10.1063/1.4726211} {\bibfield  {journal} {\bibinfo  {journal}
  {Journal of Applied Physics}\ }\textbf {\bibinfo {volume} {111}},\ \bibinfo
  {pages} {112612} (\bibinfo {year} {2012})}\BibitemShut {NoStop}%
\bibitem [{\citenamefont {Pandey}\ \emph
  {et~al.}(2014{\natexlab{a}})\citenamefont {Pandey}, \citenamefont
  {Gyanchandani}, \citenamefont {Somayazulu}, \citenamefont {Dey},
  \citenamefont {Sharma},\ and\ \citenamefont {Sikka}}]{Pandey2014a}%
  \BibitemOpen
  \bibfield  {author} {\bibinfo {author} {\bibfnamefont {K.~K.}\ \bibnamefont
  {Pandey}}, \bibinfo {author} {\bibfnamefont {J.}~\bibnamefont
  {Gyanchandani}}, \bibinfo {author} {\bibfnamefont {M.}~\bibnamefont
  {Somayazulu}}, \bibinfo {author} {\bibfnamefont {G.~K.}\ \bibnamefont {Dey}},
  \bibinfo {author} {\bibfnamefont {S.~M.}\ \bibnamefont {Sharma}}, \ and\
  \bibinfo {author} {\bibfnamefont {S.~K.}\ \bibnamefont {Sikka}},\ }\href
  {\doibase 10.1063/1.4884436} {\bibfield  {journal} {\bibinfo  {journal}
  {Journal of Applied Physics}\ }\textbf {\bibinfo {volume} {115}},\ \bibinfo
  {pages} {233513} (\bibinfo {year} {2014}{\natexlab{a}})}\BibitemShut
  {NoStop}%
\bibitem [{\citenamefont {Gyanchandani}\ \emph {et~al.}(1990)\citenamefont
  {Gyanchandani}, \citenamefont {Gupta}, \citenamefont {Sikka},\ and\
  \citenamefont {Chidambaram}}]{Gyanchandani1990}%
  \BibitemOpen
  \bibfield  {author} {\bibinfo {author} {\bibfnamefont {J.~S.}\ \bibnamefont
  {Gyanchandani}}, \bibinfo {author} {\bibfnamefont {S.~C.}\ \bibnamefont
  {Gupta}}, \bibinfo {author} {\bibfnamefont {S.~K.}\ \bibnamefont {Sikka}}, \
  and\ \bibinfo {author} {\bibfnamefont {R.}~\bibnamefont {Chidambaram}},\
  }\href {\doibase 10.1088/0953-8984/2/2/006} {\bibfield  {journal} {\bibinfo
  {journal} {Journal of Physics: Condensed Matter}\ }\textbf {\bibinfo {volume}
  {2}},\ \bibinfo {pages} {301} (\bibinfo {year} {1990})}\BibitemShut {NoStop}%
\bibitem [{\citenamefont {Ahuja}\ \emph {et~al.}(1993)\citenamefont {Ahuja},
  \citenamefont {Wills}, \citenamefont {Johansson},\ and\ \citenamefont
  {Eriksson}}]{Ahuja1993}%
  \BibitemOpen
  \bibfield  {author} {\bibinfo {author} {\bibfnamefont {R.}~\bibnamefont
  {Ahuja}}, \bibinfo {author} {\bibfnamefont {J.~M.}\ \bibnamefont {Wills}},
  \bibinfo {author} {\bibfnamefont {B.}~\bibnamefont {Johansson}}, \ and\
  \bibinfo {author} {\bibfnamefont {O.}~\bibnamefont {Eriksson}},\ }\href
  {\doibase 10.1103/PhysRevB.48.16269} {\bibfield  {journal} {\bibinfo
  {journal} {Physical Review B}\ }\textbf {\bibinfo {volume} {48}},\ \bibinfo
  {pages} {16269} (\bibinfo {year} {1993})}\BibitemShut {NoStop}%
\bibitem [{\citenamefont {Ostanin}\ and\ \citenamefont
  {Trubitsin}(2000)}]{Ostanin2000}%
  \BibitemOpen
  \bibfield  {author} {\bibinfo {author} {\bibfnamefont {S.}~\bibnamefont
  {Ostanin}}\ and\ \bibinfo {author} {\bibfnamefont {V.}~\bibnamefont
  {Trubitsin}},\ }\href {\doibase 10.1016/S0927-0256(00)00018-5} {\bibfield
  {journal} {\bibinfo  {journal} {Computational Materials Science}\ }\textbf
  {\bibinfo {volume} {17}},\ \bibinfo {pages} {174} (\bibinfo {year}
  {2000})}\BibitemShut {NoStop}%
\bibitem [{\citenamefont {Jomard}\ \emph {et~al.}(1998)\citenamefont {Jomard},
  \citenamefont {Magaud},\ and\ \citenamefont {Pasturel}}]{Jomard1998}%
  \BibitemOpen
  \bibfield  {author} {\bibinfo {author} {\bibfnamefont {G.}~\bibnamefont
  {Jomard}}, \bibinfo {author} {\bibfnamefont {L.}~\bibnamefont {Magaud}}, \
  and\ \bibinfo {author} {\bibfnamefont {A.}~\bibnamefont {Pasturel}},\
  }\href@noop {} {\bibfield  {journal} {\bibinfo  {journal} {Philos. Mag. B}\
  }\textbf {\bibinfo {volume} {77}},\ \bibinfo {pages} {67} (\bibinfo {year}
  {1998})}\BibitemShut {NoStop}%
\bibitem [{\citenamefont {Fang}\ \emph {et~al.}(2011)\citenamefont {Fang},
  \citenamefont {Gu}, \citenamefont {Liu}, \citenamefont {Liu}, \citenamefont
  {Huang}, \citenamefont {Ni}, \citenamefont {Li},\ and\ \citenamefont
  {Wang}}]{Fang2011}%
  \BibitemOpen
  \bibfield  {author} {\bibinfo {author} {\bibfnamefont {H.}~\bibnamefont
  {Fang}}, \bibinfo {author} {\bibfnamefont {M.}~\bibnamefont {Gu}}, \bibinfo
  {author} {\bibfnamefont {B.}~\bibnamefont {Liu}}, \bibinfo {author}
  {\bibfnamefont {X.}~\bibnamefont {Liu}}, \bibinfo {author} {\bibfnamefont
  {S.}~\bibnamefont {Huang}}, \bibinfo {author} {\bibfnamefont
  {C.}~\bibnamefont {Ni}}, \bibinfo {author} {\bibfnamefont {Z.}~\bibnamefont
  {Li}}, \ and\ \bibinfo {author} {\bibfnamefont {R.}~\bibnamefont {Wang}},\
  }\href {\doibase 10.1016/j.physb.2011.02.019} {\bibfield  {journal} {\bibinfo
   {journal} {Physica B: Condensed Matter}\ }\textbf {\bibinfo {volume}
  {406}},\ \bibinfo {pages} {1744} (\bibinfo {year} {2011})}\BibitemShut
  {NoStop}%
\bibitem [{\citenamefont {Hao}\ \emph {et~al.}(2011{\natexlab{a}})\citenamefont
  {Hao}, \citenamefont {Zhu}, \citenamefont {Zhang}, \citenamefont {Ren},\ and\
  \citenamefont {Qu}}]{Hao2011a}%
  \BibitemOpen
  \bibfield  {author} {\bibinfo {author} {\bibfnamefont {Y.}~\bibnamefont
  {Hao}}, \bibinfo {author} {\bibfnamefont {J.}~\bibnamefont {Zhu}}, \bibinfo
  {author} {\bibfnamefont {L.}~\bibnamefont {Zhang}}, \bibinfo {author}
  {\bibfnamefont {H.}~\bibnamefont {Ren}}, \ and\ \bibinfo {author}
  {\bibfnamefont {J.}~\bibnamefont {Qu}},\ }\href@noop {} {\bibfield  {journal}
  {\bibinfo  {journal} {Philos. Mag. Lett.}\ }\textbf {\bibinfo {volume} {91}}
  (\bibinfo {year} {2011}{\natexlab{a}})}\BibitemShut {NoStop}%
\bibitem [{\citenamefont {Qi}\ \emph {et~al.}(2016)\citenamefont {Qi},
  \citenamefont {Wang}, \citenamefont {Chen},\ and\ \citenamefont
  {Li}}]{Qi2016}%
  \BibitemOpen
  \bibfield  {author} {\bibinfo {author} {\bibfnamefont {X.}~\bibnamefont
  {Qi}}, \bibinfo {author} {\bibfnamefont {X.}~\bibnamefont {Wang}}, \bibinfo
  {author} {\bibfnamefont {T.}~\bibnamefont {Chen}}, \ and\ \bibinfo {author}
  {\bibfnamefont {B.}~\bibnamefont {Li}},\ }\href {\doibase 10.1063/1.4945106}
  {\bibfield  {journal} {\bibinfo  {journal} {Journal of Applied Physics}\
  }\textbf {\bibinfo {volume} {119}},\ \bibinfo {pages} {125109} (\bibinfo
  {year} {2016})}\BibitemShut {NoStop}%
\bibitem [{\citenamefont {Olijnyk}\ and\ \citenamefont
  {Jephcoat}(1997)}]{Olijnyk1997}%
  \BibitemOpen
  \bibfield  {author} {\bibinfo {author} {\bibfnamefont {H.}~\bibnamefont
  {Olijnyk}}\ and\ \bibinfo {author} {\bibfnamefont {A.~P.}\ \bibnamefont
  {Jephcoat}},\ }\href@noop {} {\ \textbf {\bibinfo {volume} {56}},\ \bibinfo
  {pages} {751} (\bibinfo {year} {1997})}\BibitemShut {NoStop}%
\bibitem [{\citenamefont {Ponosov}\ \emph {et~al.}(2012)\citenamefont
  {Ponosov}, \citenamefont {Streltsov},\ and\ \citenamefont
  {Syassen}}]{Ponosov2012}%
  \BibitemOpen
  \bibfield  {author} {\bibinfo {author} {\bibfnamefont {Y.~S.}\ \bibnamefont
  {Ponosov}}, \bibinfo {author} {\bibfnamefont {S.~V.}\ \bibnamefont
  {Streltsov}}, \ and\ \bibinfo {author} {\bibfnamefont {K.}~\bibnamefont
  {Syassen}},\ }\href@noop {} {\bibfield  {journal} {\bibinfo  {journal} {High
  Press.Res.}\ }\textbf {\bibinfo {volume} {32}} (\bibinfo {year}
  {2012})}\BibitemShut {NoStop}%
\bibitem [{\citenamefont {Liu}\ \emph {et~al.}(2008)\citenamefont {Liu},
  \citenamefont {Li}, \citenamefont {Wang}, \citenamefont {Zhang},\ and\
  \citenamefont {Zhao}}]{Liu2008}%
  \BibitemOpen
  \bibfield  {author} {\bibinfo {author} {\bibfnamefont {W.}~\bibnamefont
  {Liu}}, \bibinfo {author} {\bibfnamefont {B.}~\bibnamefont {Li}}, \bibinfo
  {author} {\bibfnamefont {L.}~\bibnamefont {Wang}}, \bibinfo {author}
  {\bibfnamefont {J.}~\bibnamefont {Zhang}}, \ and\ \bibinfo {author}
  {\bibfnamefont {Y.}~\bibnamefont {Zhao}},\ }\href {\doibase
  10.1063/1.2987001} {\bibfield  {journal} {\bibinfo  {journal} {Journal of
  Applied Physics}\ }\textbf {\bibinfo {volume} {104}},\ \bibinfo {pages}
  {076102} (\bibinfo {year} {2008})}\BibitemShut {NoStop}%
\bibitem [{\citenamefont {Akahama}\ and\ \citenamefont
  {Kawamura}(2006)}]{Akahama2006}%
  \BibitemOpen
  \bibfield  {author} {\bibinfo {author} {\bibfnamefont {Y.}~\bibnamefont
  {Akahama}}\ and\ \bibinfo {author} {\bibfnamefont {H.}~\bibnamefont
  {Kawamura}},\ }\href {\doibase 10.1063/1.2335683} {\bibfield  {journal}
  {\bibinfo  {journal} {Journal of Applied Physics}\ }\textbf {\bibinfo
  {volume} {100}},\ \bibinfo {pages} {043516} (\bibinfo {year}
  {2006})}\BibitemShut {NoStop}%
\bibitem [{\citenamefont {Goncharov}\ \emph {et~al.}(2002)\citenamefont
  {Goncharov}, \citenamefont {Gregoryanz}, \citenamefont {Struzhkin},
  \citenamefont {Hemley}, \citenamefont {Mao}, \citenamefont {Boctor},\ and\
  \citenamefont {Huang}}]{Goncharov2002}%
  \BibitemOpen
  \bibfield  {author} {\bibinfo {author} {\bibfnamefont {A.~F.}\ \bibnamefont
  {Goncharov}}, \bibinfo {author} {\bibfnamefont {E.}~\bibnamefont
  {Gregoryanz}}, \bibinfo {author} {\bibfnamefont {V.~V.}\ \bibnamefont
  {Struzhkin}}, \bibinfo {author} {\bibfnamefont {R.~J.}\ \bibnamefont
  {Hemley}}, \bibinfo {author} {\bibfnamefont {H.~K.}\ \bibnamefont {Mao}},
  \bibinfo {author} {\bibfnamefont {N.}~\bibnamefont {Boctor}}, \ and\ \bibinfo
  {author} {\bibfnamefont {E.}~\bibnamefont {Huang}},\ }\href@noop {}
  {\bibfield  {journal} {\bibinfo  {journal} {Proceedings of the International
  School of Physics, “Enrico Fermi” Course CXLVII, R. J. Hemley,
  G.L.Chiarotti, M. Bernasconi, and L. Ulivi}\ ,\ \bibinfo {pages} {297}}
  (\bibinfo {year} {2002})}\BibitemShut {NoStop}%
\bibitem [{\citenamefont {Pandey}\ \emph
  {et~al.}(2014{\natexlab{b}})\citenamefont {Pandey}, \citenamefont
  {Gyanchandani}, \citenamefont {Somayazulu}, \citenamefont {Dey},
  \citenamefont {Sharma},\ and\ \citenamefont {Sikka}}]{Pandey2014}%
  \BibitemOpen
  \bibfield  {author} {\bibinfo {author} {\bibfnamefont {K.~K.}\ \bibnamefont
  {Pandey}}, \bibinfo {author} {\bibfnamefont {J.}~\bibnamefont
  {Gyanchandani}}, \bibinfo {author} {\bibfnamefont {M.}~\bibnamefont
  {Somayazulu}}, \bibinfo {author} {\bibfnamefont {G.~K.}\ \bibnamefont {Dey}},
  \bibinfo {author} {\bibfnamefont {S.~M.}\ \bibnamefont {Sharma}}, \ and\
  \bibinfo {author} {\bibfnamefont {S.~K.}\ \bibnamefont {Sikka}},\ }\href
  {\doibase 10.1063/1.4884436} {\bibfield  {journal} {\bibinfo  {journal}
  {Journal of Applied Physics}\ }\textbf {\bibinfo {volume} {115}},\ \bibinfo
  {pages} {0} (\bibinfo {year} {2014}{\natexlab{b}})}\BibitemShut {NoStop}%
\bibitem [{\citenamefont {Kresse}\ and\ \citenamefont
  {Furthm{\"{u}}ller}(1996)}]{Kresse1996}%
  \BibitemOpen
  \bibfield  {author} {\bibinfo {author} {\bibfnamefont {G.}~\bibnamefont
  {Kresse}}\ and\ \bibinfo {author} {\bibfnamefont {J.}~\bibnamefont
  {Furthm{\"{u}}ller}},\ }\href {\doibase 10.1103/PhysRevB.54.11169} {\bibfield
   {journal} {\bibinfo  {journal} {Physical Review B}\ }\textbf {\bibinfo
  {volume} {54}},\ \bibinfo {pages} {11169} (\bibinfo {year}
  {1996})}\BibitemShut {NoStop}%
\bibitem [{\citenamefont {Bl{\"{o}}chl}(1994)}]{Blochl1994}%
  \BibitemOpen
  \bibfield  {author} {\bibinfo {author} {\bibfnamefont {P.~E.}\ \bibnamefont
  {Bl{\"{o}}chl}},\ }\href {\doibase 10.1103/PhysRevB.50.17953} {\bibfield
  {journal} {\bibinfo  {journal} {Physical Review B}\ }\textbf {\bibinfo
  {volume} {50}},\ \bibinfo {pages} {17953} (\bibinfo {year}
  {1994})}\BibitemShut {NoStop}%
\bibitem [{\citenamefont {Perdew}\ and\ \citenamefont
  {Burke}(1996)}]{Perdew1996}%
  \BibitemOpen
  \bibfield  {author} {\bibinfo {author} {\bibfnamefont {J.~P.}\ \bibnamefont
  {Perdew}}\ and\ \bibinfo {author} {\bibfnamefont {K.}~\bibnamefont {Burke}},\
  }\href {\doibase 10.1002/(SICI)1097-461X(1996)57:3<309::AID-QUA4>3.0.CO;2-1}
  {\bibfield  {journal} {\bibinfo  {journal} {International Journal of Quantum
  Chemistry}\ }\textbf {\bibinfo {volume} {57}},\ \bibinfo {pages} {309}
  (\bibinfo {year} {1996})}\BibitemShut {NoStop}%
\bibitem [{\citenamefont {Stassis}\ \emph {et~al.}(1981)\citenamefont
  {Stassis}, \citenamefont {Arch}, \citenamefont {McMasters},\ and\
  \citenamefont {Harmon}}]{Stassis1981}%
  \BibitemOpen
  \bibfield  {author} {\bibinfo {author} {\bibfnamefont {C.}~\bibnamefont
  {Stassis}}, \bibinfo {author} {\bibfnamefont {D.}~\bibnamefont {Arch}},
  \bibinfo {author} {\bibfnamefont {O.~D.}\ \bibnamefont {McMasters}}, \ and\
  \bibinfo {author} {\bibfnamefont {B.~N.}\ \bibnamefont {Harmon}},\ }\href
  {\doibase 10.1103/PhysRevB.24.730} {\bibfield  {journal} {\bibinfo  {journal}
  {Physical Review B}\ }\textbf {\bibinfo {volume} {24}},\ \bibinfo {pages}
  {730} (\bibinfo {year} {1981})}\BibitemShut {NoStop}%
\bibitem [{\citenamefont {Ranganathan}\ and\ \citenamefont
  {Ostoja-Starzewski}(2008)}]{Ranganathan2008}%
  \BibitemOpen
  \bibfield  {author} {\bibinfo {author} {\bibfnamefont {S.~I.}\ \bibnamefont
  {Ranganathan}}\ and\ \bibinfo {author} {\bibfnamefont {M.}~\bibnamefont
  {Ostoja-Starzewski}},\ }\href {\doibase 10.1103/PhysRevLett.101.055504}
  {\bibfield  {journal} {\bibinfo  {journal} {Physical Review Letters}\
  }\textbf {\bibinfo {volume} {101}},\ \bibinfo {pages} {3} (\bibinfo {year}
  {2008})}\BibitemShut {NoStop}%
\bibitem [{\citenamefont {Verma}\ and\ \citenamefont
  {Upadhyaya}(1971)}]{Verma1971}%
  \BibitemOpen
  \bibfield  {author} {\bibinfo {author} {\bibfnamefont {M.~P.}\ \bibnamefont
  {Verma}}\ and\ \bibinfo {author} {\bibfnamefont {J.~C.}\ \bibnamefont
  {Upadhyaya}},\ }\href {\doibase 10.1088/0305-4608/1/5/315} {\bibfield
  {journal} {\bibinfo  {journal} {Journal of Physics F: Metal Physics}\
  }\textbf {\bibinfo {volume} {1}},\ \bibinfo {pages} {315} (\bibinfo {year}
  {1971})}\BibitemShut {NoStop}%
\bibitem [{\citenamefont {Olijnyk}\ and\ \citenamefont
  {Jephcoat}(2002)}]{Olijnyk2002}%
  \BibitemOpen
  \bibfield  {author} {\bibinfo {author} {\bibfnamefont {H.}~\bibnamefont
  {Olijnyk}}\ and\ \bibinfo {author} {\bibfnamefont {A.~P.}\ \bibnamefont
  {Jephcoat}},\ }\href {\doibase 10.1007/s11661-002-1003-7} {\bibfield
  {journal} {\bibinfo  {journal} {Metallurgical and Materials Transactions A}\
  }\textbf {\bibinfo {volume} {33}},\ \bibinfo {pages} {743} (\bibinfo {year}
  {2002})}\BibitemShut {NoStop}%
\bibitem [{\citenamefont {Mouhat}\ and\ \citenamefont
  {Coudert}(2014)}]{Mouhat2014}%
  \BibitemOpen
  \bibfield  {author} {\bibinfo {author} {\bibfnamefont {F.}~\bibnamefont
  {Mouhat}}\ and\ \bibinfo {author} {\bibfnamefont {F.-X.}\ \bibnamefont
  {Coudert}},\ }\href {\doibase 10.1103/PhysRevB.90.224104} {\bibfield
  {journal} {\bibinfo  {journal} {Physical Review B}\ }\textbf {\bibinfo
  {volume} {90}},\ \bibinfo {pages} {224104} (\bibinfo {year}
  {2014})}\BibitemShut {NoStop}%
\bibitem [{\citenamefont {Hill}(1952)}]{Hill1952}%
  \BibitemOpen
  \bibfield  {author} {\bibinfo {author} {\bibfnamefont {R.}~\bibnamefont
  {Hill}},\ }\href {\doibase 10.1088/0370-1298/65/5/307} {\bibfield  {journal}
  {\bibinfo  {journal} {Proceedings of the Physical Society. Section A}\
  }\textbf {\bibinfo {volume} {65}},\ \bibinfo {pages} {349} (\bibinfo {year}
  {1952})}\BibitemShut {NoStop}%
\bibitem [{\citenamefont {Hao}\ \emph {et~al.}(2011{\natexlab{b}})\citenamefont
  {Hao}, \citenamefont {Zhu}, \citenamefont {Zhang}, \citenamefont {Ren},\ and\
  \citenamefont {Qu}}]{Hao2011}%
  \BibitemOpen
  \bibfield  {author} {\bibinfo {author} {\bibfnamefont {Y.}~\bibnamefont
  {Hao}}, \bibinfo {author} {\bibfnamefont {J.}~\bibnamefont {Zhu}}, \bibinfo
  {author} {\bibfnamefont {L.}~\bibnamefont {Zhang}}, \bibinfo {author}
  {\bibfnamefont {H.}~\bibnamefont {Ren}}, \ and\ \bibinfo {author}
  {\bibfnamefont {J.}~\bibnamefont {Qu}},\ }\href {\doibase
  10.1080/09500839.2010.529087} {\bibfield  {journal} {\bibinfo  {journal}
  {Philosophical Magazine Letters}\ }\textbf {\bibinfo {volume} {91}},\
  \bibinfo {pages} {61} (\bibinfo {year} {2011}{\natexlab{b}})}\BibitemShut
  {NoStop}%
\bibitem [{\citenamefont {de~Jong}\ \emph {et~al.}(2015)\citenamefont
  {de~Jong}, \citenamefont {Chen}, \citenamefont {Angsten}, \citenamefont
  {Jain}, \citenamefont {Notestine}, \citenamefont {Gamst}, \citenamefont
  {Sluiter}, \citenamefont {Krishna~Ande}, \citenamefont {van~der Zwaag},
  \citenamefont {Plata}, \citenamefont {Toher}, \citenamefont {Curtarolo},
  \citenamefont {Ceder}, \citenamefont {Persson},\ and\ \citenamefont
  {Asta}}]{DeJong2015}%
  \BibitemOpen
  \bibfield  {author} {\bibinfo {author} {\bibfnamefont {M.}~\bibnamefont
  {de~Jong}}, \bibinfo {author} {\bibfnamefont {W.}~\bibnamefont {Chen}},
  \bibinfo {author} {\bibfnamefont {T.}~\bibnamefont {Angsten}}, \bibinfo
  {author} {\bibfnamefont {A.}~\bibnamefont {Jain}}, \bibinfo {author}
  {\bibfnamefont {R.}~\bibnamefont {Notestine}}, \bibinfo {author}
  {\bibfnamefont {A.}~\bibnamefont {Gamst}}, \bibinfo {author} {\bibfnamefont
  {M.}~\bibnamefont {Sluiter}}, \bibinfo {author} {\bibfnamefont
  {C.}~\bibnamefont {Krishna~Ande}}, \bibinfo {author} {\bibfnamefont
  {S.}~\bibnamefont {van~der Zwaag}}, \bibinfo {author} {\bibfnamefont {J.~J.}\
  \bibnamefont {Plata}}, \bibinfo {author} {\bibfnamefont {C.}~\bibnamefont
  {Toher}}, \bibinfo {author} {\bibfnamefont {S.}~\bibnamefont {Curtarolo}},
  \bibinfo {author} {\bibfnamefont {G.}~\bibnamefont {Ceder}}, \bibinfo
  {author} {\bibfnamefont {K.~A.}\ \bibnamefont {Persson}}, \ and\ \bibinfo
  {author} {\bibfnamefont {M.}~\bibnamefont {Asta}},\ }\href {\doibase
  10.1038/sdata.2015.9} {\bibfield  {journal} {\bibinfo  {journal} {Scientific
  Data}\ }\textbf {\bibinfo {volume} {2}},\ \bibinfo {pages} {150009} (\bibinfo
  {year} {2015})}\BibitemShut {NoStop}%
\bibitem [{\citenamefont {Bashkin}\ \emph {et~al.}(2004)\citenamefont
  {Bashkin}, \citenamefont {Nefedova}, \citenamefont {Tissen},\ and\
  \citenamefont {Ponyatovsky}}]{Bashkin2004}%
  \BibitemOpen
  \bibfield  {author} {\bibinfo {author} {\bibfnamefont {I.~O.}\ \bibnamefont
  {Bashkin}}, \bibinfo {author} {\bibfnamefont {M.~V.}\ \bibnamefont
  {Nefedova}}, \bibinfo {author} {\bibfnamefont {V.~G.}\ \bibnamefont
  {Tissen}}, \ and\ \bibinfo {author} {\bibfnamefont {E.~G.}\ \bibnamefont
  {Ponyatovsky}},\ }\href {\doibase 10.1134/1.1857274} {\bibfield  {journal}
  {\bibinfo  {journal} {Journal of Experimental and Theoretical Physics
  Letters}\ }\textbf {\bibinfo {volume} {80}},\ \bibinfo {pages} {655}
  (\bibinfo {year} {2004})}\BibitemShut {NoStop}%
\end{thebibliography}%

\end{document}